# Non-Doppler Redshift and Dark Matter in the Coma Cluster

Yi-Jia Zheng

National Astronomical Observatory, Chinese Academy of Sciences, Beijing, 100012 China

**Abstract**

In 1929 Zwicky proposed a tired light theory to interpret the Hubble law (Hubble 1929). The key of the tired light theory is to interpret the observed redshift of galaxy as the non-Doppler effect. But the derivation of the dark matter in the Coma cluster proposed by Zwicky in 1933 was based on the interpretation that redshifts of galaxies were the Doppler effect, and the non-Doppler effect was not considered at all. However, if there is a reasonable non-Doppler effect and the great majority of the observed redshifts of galaxies in the Coma cluster can be interpreted by the non-Doppler effect, then it's not needed to introduce the dark matter in the Coma cluster to keep the stability of the cluster. In this paper it is shown that the great majority of the redshifts of galaxies in the Coma cluster are caused by the non-Doppler redshift proposed by Zheng (2013), and a numerical estimation of the redshifts of galaxies in the Coma cluster is presented.

**Keywords:** redshift, dark matter, non-Doppler effect, cluster of galaxy, tired light

## 1. INTRODUCTION

When Edwin Hubble in 1929 discovered a somewhat linear relationship between the distance to a galaxy and its redshift expressed as a velocity, Zwicky (1929) immediately pointed out that the redshifts of distant galaxies were not due to motions of the galaxies, but to an unknown phenomenon that caused photons to lose energy as they traveled through space.

While examining the Coma cluster in 1933, Zwicky was the first to use the Virial theorem to infer the existence of unseen matter, which he referred to as dark matter. He calculated the gravitational mass of the galaxies within the cluster and obtained a value at least 400 times greater than expected from their luminosity, which means that most of the matter in the Coma cluster must be dark. In his calculation the velocity dispersion of galaxies in the Coma cluster was derived from the observed redshifts of galaxies using the Doppler effect and the non-Doppler effect was not considered at all.

The observed redshifts of galaxies in the Coma cluster distributed in a range from 4000 to 10000 km s$^{-1}$. The estimate of the angular radius of the Coma cluster was about 3$^o$. Fig. 1(a) shows the distribution of galaxies in the Coma cluster (Gregory 1975). Assume that the redshift of the center of Coma cluster is 6900 km s$^{-1}$. The Hubble constant is 70km s$^{-1}$Mpc$^{-1}$. Then, the distance of the Coma cluster is about 100 Mpc. If the great majority of redshifts of galaxies in the cluster were caused by the expand of the universe as the Hubble law describes, then the scale of the Coma cluster along the line of sight will be about 80 Mpc. The scale perpendicular to the line of sight will be about 13 Mpc only. This is conflict with the isotropy assumption in the cosmology. Fig. 1(b) is the same as Fig. 1(a) but the scale of the line of sight has been normalized according to the Hubble law.

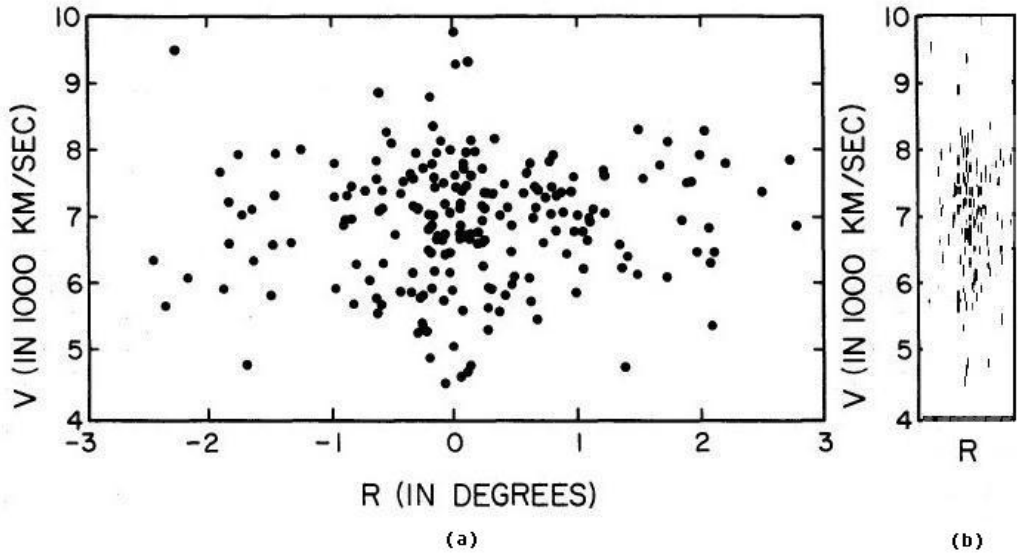

Fig.1. (**a**) The distribution of galaxies in the Coma cluster (from Gregory 1975); (**b**) same as (**a**) but the scale of the line of sight has been normalized according to the Hubble law.

If the great majority of observed redshifts of galaxies in the Coma cluster were due to the internal velocity dispersion, then according to the Virial theory, the mass derived from the luminosity of the Coma cluster was too small to keep the stability of the cluster. To stabilize the Coma cluster the dark matter approximately 7 times the mass of the galaxies must be introduced.

In contrast, if the great majority of observed redshifts of the galaxies in the Coma cluster were caused by a non-velocity effect, such as the 'soft-photon process', then the velocity dispersion of galaxies in the Come cluster could not be derived directly from the redshift dispersion of galaxies in the cluster. Therefore, it may not be needed to introduce dark matter to stabilize the Coma cluster. In the next section a quantitative estimation of the non-Doppler redshift (Zheng 2013) for the Coma cluster is presented.

## 2. QUANTITATIVE ESTIMATION

The observations show that the distribution of the galaxies in the Coma cluster and that of the X-ray gas are remarkably similar. The observations of X-ray show that in the Coma cluster the intergalactic gas is of a $10^8\,°K$ high temperature (David et al. 1993). Therefore, the intergalactic gas in the Coma cluster is almost fully ionized. The electron number density in the center of the Coma cluster is about $2.7 \times 10^{-3}\,cm^{-3}$ (Lea et al. 1973). Due to the non-Doppler redshift effect, the observed redshifts of galaxies in the Coma cluster will include the non-velocity redshifts caused by the soft photon process (Zheng 2013). Compared with the observed redshift of a galaxy located at the center of the Coma cluster, due to the non-velocity redshift, the observed redshifts of galaxies located at the front and back ends of the center will be less and larger, respectively. In Zwicky's calculation, he did not consider this non-velocity redshift effect at all. Therefore, the derived velocity dispersion by Zwicky using the redshift dispersion would be larger than the real velocity dispersion. This is why Zwicky obtained a much larger gravitational mass for the stability of the Coma cluster.

As an estimation, if the distribution of the electron number density can be determined, then the non-Doppler redshift effect can be estimated numerically. Assuming the electron number density in the center is $n_e(0)=2.5\times10^{-3}$ cm$^{-3}$ (Lea et al. 1973), taking the factor $\frac{2e^2}{\pi.\eta}\frac{\ln\Lambda}{V_e}$ equal to 2 approximately (Zheng 2013), and assuming the distribution of the electron number density obeys the Gauss law with a characteristic radius $r_o$ =1 Mp$c$, i.e. (?),

$$n_e(r) = n_e(0) \times \exp(-\frac{r^2}{r_o^2})$$

then according to equation (6) in the manuscript APJ19572, the non-Doppler redshift effect caused by the intergalactic electron contribution for the galaxy in the center of cluster is about 3000 km/s. Therefore, the great majority of the observed redshift dispersion of the galaxies in the Coma cluster can be caused by the non-velocity effect. The real velocity dispersion of the galaxies in the Coma cluster will be much smaller than the value derived by Zwicky. Taking this effect into account, the gravitational mass derived from the Virial theorem in the Coma cluster is much smaller than the value obtained by Zwicky. Consequently, it's not needed to introduce dark matter in the Coma cluster for the stability of the cluster.

## 3. DISCUSSION

If there is no non-Doppler redshift effect in the Coma cluster, then the observed redshifts of galaxies in the Coma cluster will include the cosmological redshifts (the fraction that obeys the Hubble law) and redshifts caused by internal random peculiar motion. For a galaxy located at the center of the Coma cluster, the observed redshift is approximately equal to the average value of the observed redshifts of the galaxies in the Coma cluster, and the observed redshift of the galaxy will be approximately equal to the cosmological redshift. Therefore, the redshift of the galaxy in the center of the Coma cluster in the redshift - distance Hubble diagram should be located near the straight line derived from the Hubble law. However, in the Hubble diagram the Coma cluster is not located near the straight line derived from the Hubble law (Sandage 1972). This seems to be a little suspicious.

Taking the non-velocity redshift effect into account, the observed redshift of the galaxy at the center of the Coma cluster will be larger than the redshift derived (?) from the Hubble law using approximately 3000 km/s, the real cosmological redshift will be reduced to about 4000 km/s. That is why in the Hubble diagram the Coma cluster deviates from the standard straight (?) line. Using 4000 km/s instead the 6900 km/s as the cosmological redshift of the center galaxy of the Coma cluster, the Coma cluster in the Hubble diagram will move to the position that is close to the straight (?) line derived (?) from the Hubble law (see Fig. 2).

Meanwhile, during the soft photon process, the emitted soft photons will be absorbed by the intergalactic gas. The intergalactic gas will be heated and its high temperature can be maintained. Therefore, the soft photon process is a more reasonable explanation than the dark matter for the huge redshift dispersion in the Coma cluster.

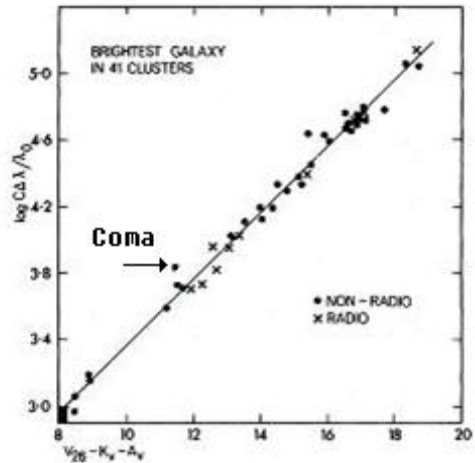

Fig. 2. the Coma cluster in the Hubble diagram (from Sandage 1972).

## Acknowledgment

I would like to thank Dr. Nailong Wu for the correction and suggestions his made to greatly improve the English of the manuscript.